\documentclass[12pt,journal,onecolumn,draftclsnofoot]{IEEEtran}
\usepackage{amsmath}
\usepackage{amssymb}
\usepackage{cite}
\usepackage{graphicx}
\usepackage{relsize}
\usepackage{multirow}
\usepackage{cancel}
\usepackage{alltt}
\usepackage{algorithm}
\usepackage{algpseudocode}
\usepackage{mathtools}

\newtheorem{theorem}{Theorem}

\newtheorem{lemma}{Lemma}
\newtheorem{example}{Example}

\begin{document}

\title{Bit Flipping Moment Balancing Schemes for Insertion, Deletion and Substitution Error Correction}
\author{Ling Cheng and Hendrik C. Ferreira %
\thanks{L. Cheng is with the School of Electrical and Information Engineering, University of the Witwatersrand, Private Bag 3, Wits. 2050, Johannesburg, South Africa. (email: ling.cheng@wits.ac.za). H.~C. Ferreira is with the Department of Electrical and Electronic Engineering Science, University of Johannesburg, Auckland Park, 2006, South Africa. (e-mail: hcferreira@uj.ac.za).}
}
\maketitle
\begin{abstract}In this paper, two moment balancing schemes, namely a variable index scheme and a fixed index scheme, for either single insertion/deletion error correction or multiple substitution error correction are introduced for coded sequences originally developed for correcting substitution errors only. By judiciously flipping bits of the original substitution error correcting code word, the resulting word is able to correct either a reduced number of substitution errors or a single insertion/deletion error. The number of flips introduced by the two schemes can be kept small compared to the code length. It shows a practical value of applying the schemes to a long substitution error correcting code for a severe channel where substitution errors dominate but insertion/deletion errors can occur with a low probability. The new schemes can be more easily implemented in an existing coding system than any previously published moment balancing templates since no additional parity bits are required which also means the code rate remains same and the existing substitution error correcting decoder requires no changes. Moreover, the work extends the class of Levenshtein codes capable of correcting either single substitution or single insertion/deletion errors to codes capable of correcting either multiple substitution errors or single insertion/deletion error.

\end{abstract}


\IEEEpeerreviewmaketitle

\section{Introduction}
\label{sec1}

Synchronization errors at symbol level are defined as insertion and deletion errors. During a transmission, the event that an unknown symbol is put in at an unknown index is called an insertion error and the event that at an unknown index an unknown symbol is left out is called a deletion error. The moment balancing template technique was investigated to correct insertion/deletion errors \cite{Ferreira09, Cheng10, Cheng12}. In this paper, we further extend this early work to two new schemes, which can correct either substitution errors or a single insertion/deletion error.

By using number theory, codes were invented to correct asymmetric errors, substitution errors and insertion/deletion errors \cite{Varshamov65, Levenshtein65b, Tenengolts76, Constantin79, Tenengolts84, Ferreira97, abdel1998detecting, khaled12}. These codes were proposed to have a deterministic code construction and a deterministic insertion/deletion error correcting capability. With `deterministic' we mean guaranteed correction of all specified error patterns, as opposed to the correction of most patterns with a high probability as in \cite{Davey01}. The difference between `deterministic' and `probabilistic' has been well addressed in \cite{khaled12}. In this paper, we focus on one construction proposed by Varshamov and Tenengolts \cite{Varshamov65}, often called the Varshamov-Tenengolts (VT) construction. Levenshtein discovered that the same construction can be used to generate codes to correct single insertion/deletion error \cite{Levenshtein65b}. The single insertion/deletion error correcting code generated based on the VT construction is called the Levenshtein code. The relationship among group theoretic codes, the VT construction and Levenshtein codes was investigated by Constantin and Rao \cite{Constantin79}. Levenshtein codes with additional rules were found to be able to correct either a single substitution error or a single insertion/deletion error \cite{Levenshtein65b}. In \cite{Tenengolts76}, a class of codes capable of correcting a deletion and a prefix substitution error was presented. The VT construction has been further implemented in single nonbinary insertion/deletion error correcting codes presented by Tenengolts \cite{Tenengolts84} and ST codes presented by Abdel-Ghaffar \cite{abdel1998detecting}. In \cite{Ferreira97}, the high-order spectrum-null code construction published in \cite{immink1987binary} was found to be a subset of a Levenstein code. Helberg and Ferreira \cite{Helberg02} presented a class of codes which can correct multiple insertion/deletion errors based on a construction which is a generalization of the Varshamov-Tenengolts construction \cite{abdel1998detecting, khaled12}. Dolecek and Anantharam \cite{Dolecek06, Dolecek07} presented a class of codes which can correct multiple repetition errors based on some high-order moment conditions. Codes for adjacent and burst insertion/deletion error correction were investigated in \cite{bours1994codes, cheng2011coding,cheng2014codes, ITA15, schoeny2016codes, smith2017interleaved}. Some recent results on multiple insertion/deletion error correcting codes can be found in \cite{ brakensiek2016efficient}. Since Schulman and Zuckerman \cite{Schulman99} presented the first asymptotically good construction of insertion/deletion error correcting codes, some recent results are presented in \cite{guruswami2016efficiently, guruswami2017deletion, haeupler2017synchronization0, haeupler2017synchronization, haeupler2017synchronization1, cheng2017synchronization, guruswami2017efficiently, guruswami2018coding, haeupler2018synchronization}. The `synchronization string' construction presented in \cite{haeupler2017synchronization0, haeupler2017synchronization, haeupler2017synchronization1, cheng2017synchronization, haeupler2018synchronization} can efficiently convert insertion/deletion errors into erasures and substitution errors by using an index en/decoding method and then the erasures and substitution errors can be further corrected.  

In this paper we further investigate two new schemes which can convert a substitution error correcting code word into a sequence which may also correct a single insertion/deletion error based on systematic encoding, the so-called moment balancing template \cite{Ferreira09}. When compared to the systematic encoding in conventional error correcting codes, there is the similarity that the parity bits have fixed indices in the sequence, while the difference is that they are not always at adjacent indices. By judiciously choosing the parity bits, the sequences generated by this scheme can correct a single insertion/deletion error. The number of parity bits is small of the same order as the number of parity bits in a Hamming code of comparable length. The idea behind this scheme is to manipulate the first-order moment property of a sequence, which is also the key to the Varshamov-Tenengolts construction and the Levenshtein codes. Moment balancing templates for different types of sequences can be found in \cite{Ferreira09, Cheng10, Cheng12}. In the previous studies on moment balancing templates, the template is composed of information bits and parity bits. It is significant that in this paper, we present two new schemes for substitution error correcting codes that require no additional parity bits to balance the moment of a sequence, thus reducing overhead and also retaining the code rate with the trade-off of possibly reducing the substitution error correcting capability to some extent. 

The main contribution of this paper is to introduce two general schemes for the moment balancing purpose, namely a variable index scheme and a fixed index scheme. Moreover, the special case of variable index schemes, the one-flip method can be considered as an extension of Levenshtein codes that can correct single insertion/deletion or single substitution error. Based on the special case, the lower bound of cardinalities of multiple substitution errors or single insertion/deletion error correcting codes can be therefore derived.

The paper is organized as follows. In Section~\ref{sec2}, we review the VT construction and the insertion/deletion/substitution error correcting codes and the moment balancing templates based on this construction. The variable index scheme as well as its special case, one-flip scheme are presented in Section~\ref{sec3}. The fixed bit flipping scheme is introduced in Section~\ref{sec4}. The analysis and discussion of the two new schemes are presented in Section~\ref{sec5}. The paper is concluded in Section~\ref{sec6}.  

\section{Varshamov-Tenengolts Construction and Moment Balancing Templates}
\label{sec2}

A brief introduction to the VT construction and some different classes of error correcting codes based on the VT construction follows.

Let ${\mathbf{x}}=x_1x_2\ldots x_n$ denote a binary code word. Given $ a \in Z_m $ for all ${\mathbf{x}} \in \mathcal{C}_a$, the moment function of ${\mathbf{x}}$ is defined as
\begin{equation}
	\sigma({\bf x})=\sum_{i=1}^{n}{ix_i} \equiv a \pmod{m}.
	\label{eq:mom}
\end{equation}

\begin{itemize}
	\item If 
	$
	\sum_{i=1}^{n}{c_ix_i} \equiv a \pmod{3},
	$
	where $c_i =1$ if $i$ is odd and $c_i =2$ if $i$ is even, $\mathcal{C}_a$ is a substitution-transposition correcting code \cite{abdel1998detecting}.
	\item If $m \geq n+1$, $\mathcal{C}_a$ is a single insertion/deletion correcting code \cite{Levenshtein65b}.
	\item If $m \geq 2n$, $\mathcal{C}_a$ is a single insertion/deletion or substitution correcting code \cite{Levenshtein65b}.
	\item If $m \geq 2n-2$ and $\sum_{i=1}^n {c_ix_i} \equiv b \pmod{2}$, where $b \in \{0, 1\}$, $\mathcal{C}_a$ is a single insertion/deletion or prefixing substitution correcting code \cite{Tenengolts76}.
\end{itemize}

Let ${\boldsymbol{\alpha}}=\alpha_1 \alpha_1 \ldots \alpha_n$ denote a binary sequence derived from ${\mathbf{x}}$ according to the relation rule
$\alpha_i = {\begin{cases} 1, \text{ if } x_i \geq x_{i-1}, \\ 
													0, \text{ if } x_i < x_{i-1}  .\\
	\end{cases}}
$ 

Here $\alpha_1$ can be any binary symbol. Tenengolts presented two selection rules to construct a non-binary single insertion/deletion correcting code as follows. When $\sum_{i=1}^n{x_i} \equiv \beta \pmod q,$ and $\sum_{i=1}^n{(i-1)\alpha_i} \equiv \gamma \pmod n,$ for some fixed integers $\beta$ and $\gamma$, $\mathcal{C}$ is a non-binary single insertion/deletion correcting code \cite{Tenengolts84}.

Note that a code that can correct $s$ deletions can also correct $s$ insertions \cite{Levenshtein65b}, where $s$ is positive integer.

A brief introduction to the moment balancing template follows.

Let $\mathcal{C}$ be a $[K,k]$ binary code, which is not necessarily linear, of length $K$ that has $2^k$ code words. Each code word $\mathbf{c} = (c_1 c_2 \cdots c_K)$ in the code $\mathcal{C}$ is mapped into a distinct sequence $\mathbf{x} = (x_1 x_2\cdots x_n)$ whose moment is congruent to a fixed integer $a$ modulo another fixed integer $m$. Similar to the systematic encoding for substitution error correcting codes, the mappings in which the code bits $c_1 c_2 \cdots c_K$ appear in fixed indices in the sequence $\mathbf{x}$, \textit{i.e.}, $c_i = x_{\gamma(i)}$ for some $1 \leq \gamma(1) < \gamma(2) < \cdots < \gamma(K) \leq n$, are achieved in a moment balancing template. The remaining bits in $\mathbf{x}$, which are called balancing bits and denoted by $b_1, b_2, \ldots, b_{n-K}$, are positioned in the $n-K$ indices that are not occupied by code bits. In particular, $b_i = x_{\beta(i)}$ where  $1\le \beta(1)<\beta(2)<\cdots<\beta(n-K)\le n$ and $\cup_{i}\gamma(i)$ and $\cup_{i}\beta(i)$ are disjoint sets whose union is $\{1, 2, \ldots, n\}$. Let $\sigma_c(\mathbf{x}) = \sum_{i=1}^{K}\gamma(i)c_i \pmod{m}$ and $\sigma_b(\mathbf{x}) = \sum_{i=1}^{n-K}\beta(i)b_i \pmod{m}$. Then, $\sigma_c(\mathbf{x})$ and $\sigma_b(\mathbf{x})$ indicate the contribution of the code bits and the balancing bits, respectively, to the moment of $\mathbf{x}$. In particular, $\sigma_c(\mathbf{x}) + \sigma_b(\mathbf{x}) \pmod{m} = \sigma(\mathbf{x}) \pmod{m}$. 

\section{Variable Index Bit Flipping Moment Balancing Scheme}
\label{sec3}
We present a moment balancing scheme by flipping bits without adding extra balancing bits to balance the moment value of a code word in order to enable a substitution error correcting code word to correct a single insertion or deletion error. By flipping a bit whose index is unknown to the receiver, we artificially create a substitution error which can be corrected together with channel errors. In our work, only the individual bits are flipped, which is different to a bit inversion operation based on the Knuth algorithm \cite{immink2004codes} for the dc-free balancing purpose, which involves a specific bit and all the following bits. In this section, we will introduce in general a variable index bit flipping moment balancing scheme, namely multiple-flip moment balancing (MFMB) scheme in Section~\ref{sec3a}, and a special variable index bit flipping moment balancing scheme, namely one-flip moment balancing (OFMB) scheme in Section~\ref{sec3b}.

\subsection{Multiple-Flip Moment Balancing Scheme}
\label{sec3a}

In the rest of the paper let $\mathcal{C}$ be a $(n,M,d_{min})$ binary code, which is not necessarily linear, of length $n$ that has $M$ code words and the minimum Hamming distance $d_{min}$. We are interested in mapping each code word $\mathbf{c} = (c_1 c_2 \cdots c_n)$ in the code $\mathcal{C}$ into a distinct binary sequence $\mathbf{x} = (x_1 x_2\cdots x_n)$ whose moment is congruent to a fixed integer modulo another fixed integer. This is possible if and only if $M$ is at most equal to the number of distinct sequences of length $n$ satisfying this congruence condition. In this paper, we focus on mappings in which by flipping a minimal number of bits in $\mathbf{c}$ to obtain $\mathbf{x}$ whose moment is congruent to a fixed integer modulo another fixed integer. Let $d_H(\cdot, \cdot)$ denote the Hamming distance of two sequences. Given $\mathbf{c} = (c_1 c_2 \cdots c_n)$ and a constant integer $a \in \{0, 1, \ldots, m-1\}$ where the constant integer $m >n$, $\mathbf{x}$ is generated on argument $\mathbf{x}=\underset{\sigma(\mathbf{x})=a}{\arg\min} d_H(\mathbf{c}, \mathbf{x})$. We further define $d$ as the maxima of all Hamming distances of each ${\mathbf{c}} \in \mathcal{C}$ and its corresponding $\bf x$, \textit{i.e.}, $d=\max_{\forall \mathbf{c} \in \mathcal{C}}  d_H(\mathbf{c}, \mathbf{x})$.

\begin{lemma}
\label{le1}
Let $\mathcal{C}$ be a $(n,M,d_{min})$ code. If $d_{min} > 2d$, all $\mathbf{x}$'s constitute a new $(n,M,d_{min}')$ single insertion/deletion error correcting code, where $d_{min}' \ge d_{min}-2d$.
\end{lemma}

\begin{IEEEproof}
Since a maximum of $d$ bit flipping operations for each code word in $\mathcal{C}$ have been carried out, any two sequences $\mathbf{x}$'s generated from two different code words in $\mathcal{C}$ are distinct given $d_{min} > 2d$. However, the minimum Hamming distance of the set of all $\mathbf{x}$'s is reduced to $d_{min}'$, where $d_{min}' \ge d_{min}-2d$.
\end{IEEEproof}

\begin{lemma}
\label{le2}
For an arbitrary sequence $\mathbf{c} = (c_1 c_2 \cdots c_n)$ of length $n$, maximally $\lfloor \log_2 n \rfloor +1$ bits need to be inverted to obtain $\mathbf{x} = (x_1 x_2\cdots x_n)$ with $\sigma(\mathbf{x})=a$, where $2^{\lfloor \log_2 n \rfloor+1}\geq m>n$.
\end{lemma}

\begin{IEEEproof}
Given the fixed indices $i \in \{2^0, 2^1, 2^2, \ldots, 2^{\lfloor \log_2 n \rfloor}\}$, when converting $\bf c$ into $\bf x$ by inverting some bits in these fixed indices, the obtained values of $\sigma(\mathbf{x})$ take on all values from 0 to $m-1$ \cite{Ferreira09}.  
\end{IEEEproof}

In Lemma~\ref{le2}, we present an upper bound of maximum number of bit flips for a code. Therefore, we have $0 \leq d \leq \lfloor \log_2 n \rfloor +1$.

The following theorem illustrates the MFMB scheme.

\begin{theorem}
Let $\mathcal{C}$ be a $(n,M,d_{min})$ code. Let $a$, $d$ and $m$ be three integers, where $0 \leq d \leq \lfloor \log_2 n \rfloor +1$, $2d < d_{min}$, $0 \leq a < m$ and $2^{\lfloor \log_2 n \rfloor+1}\geq m>n$. (a) Any code word $\bf c$ in $\mathcal C$ can be turned into a distinct sequence $\bf x$ with $\sigma(\mathbf{x})=a$ by flipping maximum $d$ bits at unknown indices. (b) All distinct sequences constitute a new code that can correct a single insertion/deletion error or at least $ \lfloor \frac{d_{min}-2d-1}{2} \rfloor$ substitution errors. (c) If all possible bit flip indices are known, the resulting code can correct a single insertion/deletion error or at least $ \lfloor \frac{d_{min}-d-1}{2} \rfloor$ substitution errors. A single insertion/deletion error correcting code which also can correct no less than $ \lfloor \frac{d_{min}-\lfloor \log_2 n \rfloor -2}{2} \rfloor$ of substitution errors is guaranteed.
\end{theorem}

\begin{IEEEproof}
First, according to Lemma~\ref{le2}, maximum $\lfloor \log_2 n \rfloor +1$ bit flips are required to satisfy the condition of $\sigma(\mathbf{x})=a$, although the possible indices of bit flips are fixed (known). The number of necessary bit flips in variable (unknown) indices cannot be more than that of fixed case. Second, since in a general case $d$ bit flips can appear at any unknown indices, according to Lemma~\ref{le1}, the resulting code can correct at least $ \lfloor \frac{d_{min}-2d-1}{2} \rfloor$ substitution errors. Furthermore, the resulting code satisfies the condition of $\sigma(\mathbf{x})=a$ and can correct a single insertion/deletion error. Third, if the possible bit flip indices are known, the errors at the unknown indices can be considered as erasures. Therefore, the resulting code can correct a single insertion/deletion error or at least $ \lfloor \frac{d_{min}-d-1}{2} \rfloor$ substitution errors. Since $d=\lfloor \log_2 n \rfloor+1$ bit flips at known indices are sufficient, the number of substitution errors can be corrected by the resulting code is lower bounded by $ \lfloor \frac{d_{min}-\lfloor \log_2 n \rfloor -2}{2} \rfloor$. 
\end{IEEEproof}

\begin{example}
For simplicity we start with a code with $d_{min}=3$ and convert it into a code correcting single insertion/deletion error. Let $a=0$. Choose $\mathcal{C}$ as a (7, 16, 3) Hamming code (the element after the code word in each row is the modulo value and the support set in each row shows the indices of inverting bits) and $m=n+1=8$. Let $S = \{ i: c_i \neq x_i \}$ be the support set to include all indices of the inverted bits in $\bf c$. It is evident that $|S|= d_H(\mathbf{c}, \mathbf{x})$.

\begin{table}[htb]
		\renewcommand{\arraystretch}{1.3}
		\caption{Bit flipping moment balancing template of a (7, 16, 3) Hamming code with $m=8$ }\label{tbl1}
		\begin{center}
		\begin{tabular}{ccccc}
				\hline
				Code word    & $\sigma(\bf{c})$  & $S$    \\
				\hline
	 				 0     0     0     0     0     0     0 &0&\{\}\\
	   			 1     0     0     1     1     1     0 &0&\{\}\\
			     1     0     1     1     0     0     0 &0&\{\}\\
 \textbf{1     1     0     0     0     1     0} &1& \{1\} or \{7\}\\
			     1     0     1     0     0     1     1 &1&\{1\}\\
\underline{0     0     0     1     0     1     1} &1&\{2, 5\}\\
           1     1     1     0     1     0     0 &3&\{3\}\\
\underline{0     1     0     1     1     0     0} &3&\{6, 7\}\\
			     0     0     1     1     1     0     1 &3&\{3\}\\
			     0     1     0     0     1     1     1 &4&\{4\}\\
			     0     1     1     0     0     0     1 &4&\{4\}\\
			     1     1     1     1     1     1     1 &4&\{4\}\\
\textbf{ 1     0     0     0     1     0     1} &5&\{3\} or \{5\}\\
\textbf{ 0     0     1     0     1     1     0} &6&\{2\} or \{6\}\\
\underline{1     1     0     1     0     0     1} &6&\{4, 6\}\\
			     0     1     1     1     0     1     0 &7&\{1\}\\				
				\hline
		\end{tabular}
		\end{center}
\end{table}

To balance the moment value of each code word in Table~\ref{tbl1} to be 0, no inverting operation is required for the first three code words. To balance code word 1100010, the first or the seventh bit is inverted. Two inverting operations are required for three code words (underlined). Since $d_{min}=3$, in a general case, two distinct code words can be flipped into one identical sequence by more than one inverting operation for either code word. Therefore, two bit flips choices will not be considered in this case.

However, it is observed that the balancing choice of a given code word is not unique. As shown in the rows where the code word are highlighted in bold, there are at least two options to balance one code word. In this example, a single insertion/deletion error correcting code is achieved as shown in Table~\ref{tbl2} by excluding the code words which require two bits to balance and therefore drain the substitution error correcting capability of the original code, and including multiple balanced code words derived from one original code word. In this case, the code rate is not comprised by using Table~\ref{tbl2} to encode. Including multiple balanced code words generated from the same original code word, however, heavily affects the substitution error correcting capability of the resulting code. The intention of showing Table~\ref{tbl2} is to demonstrate a bit flipping approach to implement the VT construction.

\begin{table}[hbt]
		\renewcommand{\arraystretch}{1.3}
		\caption{Bit-inverting moment balancing template of a (7, 16, 3) Hamming code with $m=8$}\label{tbl2}  
		\begin{center}
		\begin{tabular}{ccccc}
				\hline
				Code word    & $\sigma(\bf{c})$  & $S$    \\
				\hline
	 				 0     0     0     0     0     0     0 &0&\{\}\\
	   			 1     0     0     1     1     1     0 &0&\{\}\\
			     1     0     1     1     0     0     0 &0&\{\}\\
 \textbf{1     1     0     0     0     1     0} &1& \{1\} \\
 \textbf{1     1     0     0     0     1     0} &1& \{7\}\\
			     1     0     1     0     0     1     1 &1&\{1\}\\
           1     1     1     0     1     0     0 &3&\{3\}\\
			     0     0     1     1     1     0     1 &3&\{3\}\\
			     0     1     0     0     1     1     1 &4&\{4\}\\
			     0     1     1     0     0     0     1 &4&\{4\}\\
			     1     1     1     1     1     1     1 &4&\{4\}\\
\textbf{ 1     0     0     0     1     0     1} &5&\{3\}\\
\textbf{ 1     0     0     0     1     0     1} &5&\{5\}\\
\textbf{ 0     0     1     0     1     1     0} &6&\{2\}\\
\textbf{ 0     0     1     0     1     1     0} &6&\{6\}\\
		     0     1     1     1     0     1     0 &7&\{1\}\\				
				\hline
		\end{tabular}
		\end{center}
\end{table}

To this end, we can derive the following code based on the bit flipping scheme:
\begin{equation}
\begin{Bmatrix}
0000000,1001110,1011000,0100010\\
1100011,0010011,1100100,0001101\\
0101111,0111001,1110111,1010101\\
1000001,0110110,0010100,1111010
\end{Bmatrix}.
\end{equation}

\end{example}

There is a small observation leading to the following lemma.
\begin{lemma}
\label{special_mom_bal}
Let $\sigma'=a-\sigma({\bf{c}})\pmod{m}$. If $\sigma'=i$ and $m = 2i$ , where $i \in \{1, 2, \ldots, n\}$, in order to balance the sequence $\bf c$ to have $\sigma({\bf x})= a$, only 1 bit flip at the $i$'th index is required. 
\end{lemma}

\begin{IEEEproof}
When inverting the $i$'th bit of $\bf{c}$ from 0 to 1, $\sigma'=i$. When inverting the $i$'th bit of $\bf{c}$ from 1 to 0, $\sigma'=m-i$. Let $i=m-i$, we obtain $m=2i$. Therefore, one bit flip at the $i$'th index is sufficient to obtain $\bf{x}$ from $\bf{c}$ to have $\sigma({\bf x})= a$.
\end{IEEEproof}

The special balancing case presented by Lemma~\ref{special_mom_bal} is not rare and the numerical examples can be found in Table~\ref{tbl1} for the original code words 0100111, 0110001 and 1111111 to be balanced.

Note that in the earlier example, we conceptually choose a short Hamming code with $d_{min}=3$. In practical systems, we may choose long BCH codes with larger $d_{min}$ to retain most of the substitution error correcting capability and add an insertion/deletion error correcting capability to the sequences.

The validation and efficiency (in terms of the number of bit-flips introduced) of variable index scheme depend on ($n$, $a$, $m$) and the original code. For example, it is impossible to balance the sequence 010101 for $a=0$ and $m=7$ by less than three bit-flips. Therefore, in the code construction stage, a proper selection procedure is required, which involves selecting $a$ and $m$, and/or expunging some code words to optimize the code rate and/or the error correcting capability. In the next section, we will provide a guaranteed scheme by expunging some code words.

\subsection{One-Flip Moment Balancing Scheme}
\label{sec3b}

The property presented by the following lemma is the key to the OFMB scheme.
\begin{lemma}
\label{le4}
Let $\bf c$ denote a sequence of length $n$. The $n+1$ sequences including the original sequence $\bf c$ and $n$ different sequences each have a bit-flip from $\bf c$, have at least $\lceil \frac{n}{2} \rceil +1$ different moment values in a residue system defined by \eqref{eq:mom} with modulo $m>n$.
\end{lemma}

\begin{IEEEproof}
By flipping one bit of $\bf c$ in $n$ different indices, $n$ different sequences are generated and each is different from $\bf c$. A bit-flip introduces a difference in \eqref{eq:mom} since $m>n$. There are only two types of bit-flips. Either a bit with value 0 is substituted by 1, or value 1 substituted by 0. The differences introduced to \eqref{eq:mom} by  all possible 0-to-1 (or 1-to-0) flips are all different also thanks to $m>n$. Since there are in total $n$ possible bit-flips. At least half of them are either 0-to-1 or 1-to-0 flips. Therefore, by one bit-flips at each index, at least $\lceil \frac{n}{2} \rceil $ new moment values are introduced. Including the moment value of $\bf c$, there are $\lceil \frac{n}{2} \rceil +1$ different values introduced by $\bf c$ and the sequences with one bit-flip from $\bf c$.   
\end{IEEEproof}

We consider to moment-balance a code to a new code, in which each code word has identical moment value in the residue system modulo $m=n+1$. Given a binary ($n$, $M$, $d_{min} \geq 3$) code $\mathcal C$, a new code $\mathcal C'$ can be generated by using the OFMB scheme, actually an expunging process illustrated by the following steps:
\begin{enumerate}
\item By flipping only one bit in each index of $\bf c \in \mathcal C$, $n+1$ different sequences including the original word are generated from $\bf c$. These $n+1$ sequences carry at least $\lceil \frac{n}{2} \rceil +1$ different moment values according to Lemma~\ref{le4}. For each generated moment values, we only select one sequence even if there are multiple sequences carry the same moment value. 
\item By applying Step~1 to all $\bf c \in \mathcal C$, at least $ M(\lceil \frac{n}{2} \rceil +1)$ different sequences are generated since the original code $\mathcal C$ has $d_{min} \geq 3$ and for each $\bf c$ at least $\lceil \frac{n}{2} \rceil +1$ new sequences are generated in the previous step.
\item The different sequences generated in the previous step are partitioned into $m=n+1$ sets according to \eqref{eq:mom}. Among them, the one with the biggest cardinality is chosen as the new code $\mathcal C'$.
\end{enumerate}

\begin{theorem}
\label{th2}
Let $\mathcal{C}$ be a $(n,M,d_{min} \geq 3)$ code. An $(n, M', d_{min}'\geq d_{min}-2)$ code $\mathcal C'$ exists where $M' \geq \Bigl\lceil \frac{M(\lceil \frac{n}{2} \rceil +1)}{n+1}\Bigr\rceil \geq \Bigl\lceil \frac{M}{2} \Bigr\rceil$. All code words in $\mathcal C'$ satisfy the residue system defined in \eqref{eq:mom} with $m=n+1$ and some $a$, and therefore $\mathcal C'$ is also a single insertion/deletion error correcting code.
\end{theorem}

\begin{IEEEproof}
As discussed in the last section, among the original code word and the new sequences generated from it by one flip, if there are multiple sequences carrying the same moment value only one sequence should be chosen in order to minimize the decrease of minimum Hamming distance of the resulting code. Therefore, in the steps shown earlier in this section, although $M(n+1)$ different sequences can be generated based on $\mathcal C$ and one-flip operations, only no less than $ M(\lceil \frac{n}{2} \rceil +1)$ sequences are chosen. Since the sequences are partitioned into $m=n+1$ sets, the cardinality  $M'$ of the resulting code $\mathcal C'$ satisfies
\begin{equation}
M' \geq \Bigl\lceil \frac{M(\lceil \frac{n}{2} \rceil +1)}{n+1}\Bigr\rceil \geq \Bigl\lceil \frac{M}{2} \Bigr\rceil.
\end{equation}
Hence, $\mathcal C'$ is a single insertion/deletion error correcting code. Moreover, in $\mathcal C'$ no two code words are generated through one-flip operations from the same original code word. Therefore, the minimum Hamming distance $d_{min}'$ of $\mathcal C'$ satisfies

\begin{equation}
d_{min}'\geq d_{min}-2.
\end{equation}

\end{IEEEproof}

The brute-force method as shown in the steps presented earlier in this section, which is similar to the method to implement the MFMB scheme, can be used to choose a code and an encoding table.
\section{Fixed Index Bit Flipping Moment Balancing Scheme}
\label{sec4}
We further present a fixed index bit flipping scheme in this section. According to Lemma~\ref{le2}, it is guaranteed that by flipping some bits in the fixed indices $i \in \{2^0, 2^1, 2^2, \ldots, 2^{\lfloor \log_2 n \rfloor}\}$, the moment value of the obtained sequence can be balanced. Since the indices of possible inversions are fixed, the bits in these indices can be considered as erasures at the decoder and the number of erasures is $\lfloor \log_2 n \rfloor+1$. 

An example of a fixed index bit flipping scheme follows. 

\begin{example}

Choose $\mathcal{C}$ as a (15, 32, 7) binary primitive BCH code, $a=0$ and $m=n+1=16$.

\begin{table*}[htb]
		\renewcommand{\arraystretch}{1.2}
		\caption{Fixed index bit flipping moment balancing scheme of an (15, 32, 7) BCH code with $m=16$}\label{tbl3}
		\begin{center}
		\begin{tabular}{ccccc}
				\hline
				Code word    & $\sigma(\bf{c})$  & Variable Indices Bit Flipping Code I   & Fixed Indices  Bit Flipping Code II    \\
				\hline
000000000000000&0& 000000000000000& 000000000000000 \\
100001010011011&3& 100001010011{\bf{1}}11& {\bf 01}0{\bf 1}010{\bf 0}0011011\\
010001111010110&6& 01000{\bf{0}}111010110& 0{\bf 0}0{\bf 1}011{\bf 0}1010110\\
110000101001101&11&1100{\bf 1}0101001101& {\bf 00}00001{\bf 1}1001101 \\
001000111101011&14&0{\bf 1}1000111101011& 0{\bf 1}1000111101011\\
101001101110000&15&10{\bf 01}01101110000& {\bf 01}1001101110000\\
011001000111101&8&0110010{\bf 1}0111101& 0110010{\bf 1}0111101\\
111000010100110&3&11{\bf 0}000010100110& {\bf 00}1000010100110\\
000101001101110&4&000{\bf 0}01001101110& 000{\bf 0}01001101110\\
100100011110101&7&1{\bf 1}010001{\bf 0}110101& {\bf 01}01000{\bf 0}1110101\\
010100110111000&6&{\bf 1}10100{\bf 0}10111000& 0{\bf 0}0{\bf 0}00110111000\\
110101100100011&11&1101{\bf 1}1100100011& {\bf 00}01011{\bf 1}0100011\\
001101110000101&8&0011011{\bf 0}0000101& 0011011{\bf 0}0000101\\
101100100011110&1&{\bf 0}01100100011110& {\bf 0}01100100011110\\
011100001010011&10&01110{\bf 1}001010011& 0{\bf 0}11000{\bf 1}1010011\\
111101011001000&13&1111010{\bf 0}10{\bf 1}1000& {\bf 0}11{\bf 0}010{\bf 0}1001000\\
000010100110111&11&0000101001{\bf 0}0111&{\bf 1}00{\bf 1}10100110111\\
100011110101100&14&1{\bf 1}0011110101100&1{\bf 1}0011110101100\\
010011011100001&7&{\bf 1}100110{\bf 0}1100001&{\bf 1}100110{\bf 0}1100001\\
110010001111010&0&110010001111010&110010001111010\\
001010011011100&13&001010011011{\bf 0}00& {\bf 11}1010011011100\\
101011001000111&2&101{\bf 1}1{\bf 0}001000111&1{\bf 1}1{\bf 1}110{\bf 1}1000111\\
011011100001010&1&01101110000101{\bf 1}&{\bf 10}1011100001010\\
111010110010001&4&11101011001{\bf 1}001& 111{\bf 1}101{\bf 0}0010001\\
000111101011001&5&0001{\bf 0}1101011001& {\bf 11}01111{\bf 1}1011001\\
100110111000010&0&100110111000010&100110111000010\\
010110010001111&9&010110{\bf 1}10001111&{\bf 10}01100{\bf 0}0001111\\
110111000010100&10&{\bf 0}10111{\bf 1}00010100& 1{\bf 0}01110{\bf 1}0010100\\
001111010110010&13&001{\bf 0}11{\bf 1}10110010& {\bf 11}1111010110010\\
101110000101001&2&1011100001010{\bf 1}1& 1{\bf 1}1{\bf 0}10000101001\\
011110101100100&5&0111{\bf 0}0101100100& {\bf 10}1{\bf 0}10101100100\\
111111111111111&8&1111111{\bf 0}1111111& 1111111{\bf 0}1111111\\
				\hline
		\end{tabular}
		\end{center}
\end{table*}

\end{example}

In Table~\ref{tbl3}, two obtained codes by implementing the variable index moment balancing scheme and fixed index moment balancing scheme are presented in the 3rd column and 4th column respectively. The inversions of code words in Code I are highlighted in bold. The fixed indices of Code II are the 1st, 2nd, 4th and 8th indices. Only bits in these indices are possibly inverted.   

It is evident that Code I obtained by implementing the variable index scheme, compromises the substitution error correcting capability to achieve single insertion/deletion error correction. The resulting code can also correct one substitution error. The cardinality of Code I can be further increased by including more bit-flip options. The trade-off is the substitution error correcting capability of the resulting code will be further compromised. Code II obtained by implementing fixed index scheme has the same cardinality as the original code. To decode Code II, we can consider the bits at the fixed indices are erasures. In this sense, Code II can correct a single insertion/deletion error or one substitution error in addition to four erasures.  

The following theorem illustrates the fixed index bit flipping moment balancing scheme.

\begin{theorem}
Let $\mathcal C$ be a ($n$, $M$, $d_{min}$) substitution error correcting code. Let $a$, $d$ and $m$ be three integers, where $d = \lfloor \log_2 n \rfloor +1$, $0 \leq a < m$ and $2^{\lfloor \log_2 n \rfloor+1}\geq m>n$. Any code word $\bf c$ in $\mathcal C$ can be turned into a sequence $\bf x$ with $\sigma(\mathbf{x})=a$ that can correct a single insertion/deletion error or $\lfloor \frac{d_{min}-d-1}{2} \rfloor$ substitution errors by flipping maximum $\lfloor \log_2 n \rfloor +1$ bits in the fixed indices $\{2^0, 2^1, 2^2, \ldots, 2^{\lfloor \log_2 n \rfloor}\}$ of $\bf c$.   
\end{theorem}

\begin{IEEEproof}
According to Lemma~\ref{le2}, by changing the values of the bits in the fixed indices, the moment value of these bits can take on any value between 0 to $2^{\lfloor \log_2 n \rfloor+1}-1$. Therefore, it is sufficient to turn $\bf c$ into $\bf x$, which has $\sigma(\mathbf{x})=a$. Since these bits used to balance the moment value are in the fixed indices, they can be considered as erasures by the decoder. Hence, the code can correct $\lfloor \frac{d_{min}-d-1}{2} \rfloor$ substitution errors.
\end{IEEEproof}

The decoding process for both fixed and variable index schemes can be described as follows. At the receiver, based on marker or special synchronization words inserted between frames, insertion or deletion errors are first detected. If a single insertion or deletion is detected, it can be decoded by using the algorithm presented in \cite{Levenshtein65b}. If there is no insertion or deletion error, the decoder proceeds with the procedure of substitution error correction. While correcting substitution errors in the sequences encoded by the fixed scheme, the bits in the fixed indices should be marked as erasures first. 

\section{Analysis}
\label{sec5}

Let $C(n, d_{min}, s)$ denote a code of length $n$ which has $d_{min}$ minimum Hamming distance and also can correct $s$ insertion/deletion errors. In \cite{Levenshtein65b}, Levenshtein introduces a class of codes which can correct single insertion/deletion or single substitution error $C(n,3,1)$. Equivalently, it gives a lower bound of cardinalities of $C(n,3,1)$, which is $\frac{2^n}{2n}$. In this work, we extend the code to $C(n,d_{min},1)$ and the lower bound can be further considered in the light of the schemes presented in this paper. 

For the class of codes constructed in \cite{Levenshtein65b}, the construction starts by using VT construction and the single substitution error correcting capability is a by-product. In the work, we start with a multiple substitution error correcting code and make it correct a single insertion/deletion error with a limited compromised substitution error correcting capability compared with the original code. The construction takes two steps and is deterministic.

As well known, binary MDS codes are trivial \cite{macwiliams77}. While $n$ is small, very often the performance of a linear code deteriorates drastically if its valuable substitution error correcting capability is compromised for a single insertion/deletion error correction. Therefore, if $n$ is small, non-linear codes can be considered since the new schemes are not limited to implementing linear codes. Since at most $\lfloor \log_2 n \rfloor +1$ bits are required to turn a substitution error correcting code into a single insertion/deletion error correcting code, while $n$ is large the performance of a substitution error correcting code does not degrade as much as short codes. In this case, the linear codes are preferable considering the encoding and decoding complexities.   

Here we present a new lower bound of cardinalities for a $\mathcal C(n,d_{min},1)$ code.

\begin{theorem}
\label{th4}
There always exists a code $\mathcal C(n,d_{min},1)^*$ with the cardinality $|\mathcal C(n,d_{min},1)^*| \geq \frac{2^{n-1}}{V_2(n, d_{min}+1)}$, where $V_2(n,d_{min}+1)=\sum_{i=0}^{d_{min}+1} \binom{n}{i}$. 
\end{theorem}

\begin{IEEEproof}
The Gilbert-Varsharmov (GV) bound \cite{gilbert1952comparison,varshamov1957estimate} ensures the existence of binary code $\mathcal C$ of length $n$ with the minimum Hamming distance $d_{min}$, having $|\mathcal C| \geq \frac{2^n}{V_2(n, d_{min}-1)}$. Based on this result, we first start with a binary code achieving the GV bound with the minimum Hamming distance $d_{min}+2$. This binary code has the cardinality no less than $\frac{2^{n}}{V_2(n, d_{min}+1)}$. By using the OFMB scheme, the resulting code $\mathcal C^*$ has a reduced minimum Hamming distance $d_{min}$ and a reduced cardinality no less than approximately half of the original code.
\end{IEEEproof}

To this end, we can further develop a tighter lower cardinality bound for $C(n, d_{min}, 1)$ codes as follows.
\begin{theorem}
\label{th5}
There always exists a code $C(n,d_{min},1)^*$ with the cardinality 
\begin{equation}\label{eq:th5}
\begin{split}
 &|C(n,d_{min},1)^*| \geq \max\{ \frac{2^{n-1}}{V_2(n, d_{min}+1)},\\ & 
 \frac{2^{n-\lfloor \log_2 n \rfloor - 1}}{V_2(n-\lfloor \log_2 n \rfloor -1, d_{min}-1)} \}.
\end{split}
\end{equation}
\end{theorem}

\begin{IEEEproof}
Apply the moment balancing template (MBT) to a binary code of length $n - \lfloor \log_2 n \rfloor -1$, achieving the GV bound with the minimum Hamming distance $d_{min}$. The resulting code of length $n$ has the minimum Hamming distance no less than $d_{min}$ and the cardinality no less than $ \frac{2^{n-\lfloor \log_2 n \rfloor - 1}}{V_2(n-\lfloor \log_2 n \rfloor -1, d_{min}-1)}$. Therefore, we can combine this result with the lower bound derived in Theorem~\ref{th4} and give a tighter lower bound.
\end{IEEEproof}

Levenshtein \cite{Levenshtein65b} found a class of codes which can correct a single insertion/deletion error or a substitution error ($d_{min} \geq 3$) based on the VT construction while $m \geq 2n$. Therefore, the resulting code has a cardinality no less than $\frac{2^n}{2n}$. To this end, we compare the new lower bound in Theorem~\ref{th4} with Levenshtein's result for $d_{min}=3$. Since the denominator in the new lower bound is $V_2(n, d_{min}+1)= \binom{n}{4}+ \binom{n}{3}+\binom{n}{2}+\binom{n}{1}+1$ when $d_{min}=3$, the new lower bound is not superior to Levenshtein's result. Note that Levenshtein's result considers arbitrary sequences and $d_{min}=3$ only. The new result considers a substitution error correcting code and any minimum Hamming distance. The new result not only increases the minimum Hamming distance range of the resulting codes, but also introduces a possible complexity reduction to the encoding and decoding process.  

In Table~\ref{tbl4}, we compare the code word lengths, information lengths, insertion/deletion and substitution error correcting capabilities of the OFMB scheme and the MBT scheme applied to a substitution error correcting code.

\begin{table*}[htb]
		\renewcommand{\arraystretch}{1.3}
		\caption{Comparison between OFMB scheme and moment balancing template \cite{Ferreira09} for ($n$, $k$, $d_{min}$) substitution error correcting codes.}\label{tbl4}
		\begin{center}
		\begin{tabular}{ccc}
				\hline
				   & Moment Balancing Template \cite{Ferreira09}  & One-Flip Moment Balancing Scheme    \\
				\hline
				Code word length & $n+\lfloor \log_2n \rfloor +1$ & $n$ \\
				Information length & $k$ & $k-1$ \\
				$s$ insertion/deletion correction & 1 & 1\\
				Minimum Hamming distance & $d_{min}$ & $d_{min}-2$\\ 
				\hline
		\end{tabular}
		\end{center}
\end{table*}

\begin{figure}[h]
\centering
\includegraphics[width=1\linewidth]{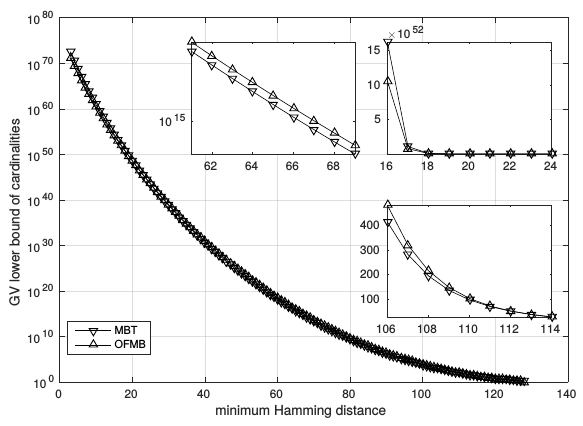}
\caption{Comparison of lower bounds of cardinalities between MBT template and OFMB scheme when the original code for the MBT template has length $256$.}
\label{comp1}
\end{figure}

We further compare the OFMB scheme with an alternative scheme, namely the MBT scheme \cite{Ferreira09}, which starts with a multiple substitution error correcting code and encodes each code word with a systematic VT construction. To implement the MBT scheme for a substitution error correcting code achieving the GV bound with $d_{min}$ of length $n-\lfloor \log_2 n \rfloor -1$, we insert $\lfloor \log_2 n \rfloor +1$ balancing bits to hold the fixed indices $\{2^0, 2^1, 2^2, \ldots, 2^{\lfloor \log_2 n \rfloor}\}$ in each code word and ensure \eqref{eq:mom} to be met by judicially choosing the values of the balancing bits.  In Fig.~\ref{comp1} we compare the cardinalities of codes  generated by the MBT template and the OFMB scheme respectively. Both resulting codes have length 265 and the original code for the MBT template has length 256. Note that both original codes are codes that achieve the GV bound. As shown in Fig.~\ref{comp1}, while the minimum Hamming distance of the resulting code takes the value between 20 to 110， the cardinality of the code generated based on the OFMB scheme is superior to the one generated by the MBT scheme.

By the following theorem we present a comparison in an asymptotic form between the lower bounds derived based on the OFMB and the MBT schemes respectively. Let $ H_2(\cdot) $ denote the binary entropy function.

\begin{theorem}[Asymptotic bound]
\label{th6}
There always exists a code $C(n,d_{min},1)^*$ with the cardinality 
\begin{equation}\label{eq:th5}
|C(n,d_{min},1)^*| \geq  2^{n-1-H_2(\frac{d_{min}+1}{n})n},
\end{equation}
while $n$ is large and $\frac{(d_{min}+1)(\lfloor \log_2 n \rfloor+1)}{2n} > 1$.
\end{theorem}

\begin{IEEEproof}
Based on Theorem~\ref{th5}, the lower bound of cardinalities is the maximum value between the lower bound derived based on the OFMB shceme, which is $\frac{2^{n-1}}{V_2(n, d_{min}+1)}$ and the one based on the MBT scheme, which is $\frac{2^{n-\lfloor \log_2 n \rfloor - 1}}{V_2(n-\lfloor \log_2 n \rfloor -1, d_{min}-1)}$. Let $\delta_1 = \frac{d_{min}+1}{n}$ and $\delta_2 = \frac{d_{min}-1}{n-\lfloor \log_2 n \rfloor -1}$. 

Since in an asymptotic form 
\begin{equation}
2^{\left(H_2(\delta)+o(1)\right)n} \leq V_2(n, \delta n) \leq 2^{H_2(\delta)n},
\end{equation}
where $0 \leq \delta  \leq \frac{1}{2}$, we can give an estimate of $  V_2(n, \delta n) = 2^{H_2(\delta)n}$. Therefore, we have 
\begin{equation}
\label{ofmb-b}
\frac{2^{n-1}}{V_2(n, d_{min}+1)} = 2^{n-1-H_2(\delta_1)n}
\end{equation} 
and
\begin{equation}
\label{mbt-b}
\frac{2^{n-\lfloor \log_2 n \rfloor - 1}}{V_2(n-\lfloor \log_2 n \rfloor -1, d_{min}-1)}=2^{n-\lfloor \log_2 n \rfloor - 1-H_2(\delta_2)(n-\lfloor \log_2 n \rfloor - 1)}.
\end{equation}

Let $ \Delta $ denote the difference between the exponents of the right terms in \eqref{ofmb-b} and \eqref{mbt-b}. We have
\begin{equation}
\begin{split}
\Delta= &\underbracket[0.8pt]{n-1-H_2(\delta_1)n}_\text{exponent of the right term in \eqref{ofmb-b}}-\left(\underbracket[0.8pt]{n-\lfloor \log_2 n \rfloor - 1-H_2(\delta_2)(n-\lfloor \log_2 n \rfloor - 1)}_\text{exponent of the right term in \eqref{mbt-b}}\right) \\
  & = \underbracket[0.8pt]{n\left(H_2(\delta_2)-H_2(\delta_1)\right)}_\text{first term} + \underbracket[0.8pt]{\lfloor \log_2 n \rfloor\left(1-H_2(\delta_2)\right)}_\text{second term} -
  \underbracket[0.8pt]{H_2(\delta_2)}_\text{third term}.
\end{split}
\end{equation}
We have the following observations:
\begin{itemize}
\item While $ n $ is large, the third term can be ignored.
\item Since $0 \leq H_2(\delta_2)\leq 1$ the second term is no less than zero.
\item Since $0 \leq \delta_1, \delta_2 \leq \frac{1}{2}$, if $\delta_2 > \delta_1$ the first term is positive.
\end{itemize}
Based on the definitions of $ \delta_1 $ and $  \delta_2 $, if $\delta_2 > \delta_1$, which means  
\begin{equation}
\frac{d_{min}-1}{n-\lfloor \log_2 n \rfloor -1} > \frac{d_{min}+1}{n},  
\end{equation}
the condition
\begin{equation}
\label{ofmb-cond}
\frac{(d_{min}+1)(\lfloor \log_2 n \rfloor+1)}{2n} > 1
\end{equation}
is required.
\end{IEEEproof}

It is evident that the new lower bound based on the OFMB scheme is guaranteed to be superior to the one derived based on the original MBT scheme if \eqref{ofmb-cond} is met, and there are a wide range of 2-tuple ($ d_{min} $ , $ n $) satisfying \eqref{ofmb-cond}.

\section{Conclusion and Future Work}
\label{sec6}

The two bit flipping schemes have three major advantages compared to the original template \cite{Ferreira09}. First, insertion/deletion errors and substitution errors are channel errors. Both should be considered while designing a code for a harsh channel, and the preferred original code is likely to be substitution error correcting code already. Second, we start with the most widely used channel codes, substitution error correcting codes. The original error correcting capability, the remaining substitution error correcting capability, and single insertion/deletion correcting capability can be balanced by using the new schemes. Since the new schemes only invert the bits if necessary to satisfy moment constraint and every flip in general will cause more deduction on the substitution error correcting capability, therefore we can say the capability can be reduced if required. Third, most modern systems already are designed based on a given substitution error correcting code. The new schemes are practical to implement on top of an existing system since they require no code length change.  

In this paper, we present an approach to reduce the capability of a substitution error correcting code to also correct a single insertion/deletion error - in fact thus a rate $R=1$ moment balancing template. The encoding and decoding procedures are discussed and the performance is evaluated.  

The efficiency of MFMB scheme can be further evaluated in future work. An analytical approach by using generation functions involving polynomial multiplication was considered by the authors to investigate how many different moment values can be generated by multiple bit-flips from a given code word. However, as shown in \cite{erdos1980old,schinzel2009number}, the analytical model of number of terms of power of polynomial is still an open question to the authors' best knowledge.

\bibliographystyle{IEEEtran}
\bibliography{bib/IEEEfull,bib/IEEEabrv,bib/mybib}

\end{document}